\documentclass[a4paper,fleqn,usenatbib]{mnras}
\usepackage{newtxtext,newtxmath}
\usepackage[T1]{fontenc}
\usepackage{ae,aecompl}
\usepackage{graphicx}
\usepackage{amsmath}
\usepackage{amssymb}

\title[Fallback accretion in LGRBs with flares]{Fallback accretion on to a newborn magnetar: long GRBs with giant X-ray flares}

\author[S. L. Gibson et al.]{
S. L. Gibson,$^{1}$\thanks{E-mail: slg44@leicester.ac.uk}
G. A. Wynn,$^{1}$
B. P. Gompertz$^{2,3}$
and P. T. O'Brien$^{1}$
\\
$^{1}$Department of Physics and Astronomy, University of Leicester, University Road, Leicester LE1 7RH\\
$^{2}$Space Telescope Science Institute, 3700 San Martin Drive, Baltimore, MD 21218, USA\\
$^{3}$Department of Physics, University of Warwick, Coventry CV4 7AL
}

\date{Accepted XXX. Received YYY; in original form ZZZ}

\pubyear{2017}
 
\begin{document}
\label{firstpage}
\pagerange{\pageref{firstpage}--\pageref{lastpage}}
\maketitle

\begin{abstract}
Flares in the X-ray afterglow of gamma-ray bursts (GRBs) share more characteristics with the prompt emission than the afterglow, such as pulse profile and contained fluence. As a result, they are believed to originate from late-time activity of the central engine and can be used to constrain the overall energy budget. In this paper, we collect a sample of $19$ long GRBs observed by \emph{Swift}-XRT that contain giant flares in their X-ray afterglows. We fit this sample with a version of the magnetar propeller model, modified to include fallback accretion. This model has already successfully reproduced extended emission in short GRBs. Our best fits provide a reasonable morphological match to the light curves. However, $16$ out of $19$ of the fits require efficiencies for the propeller mechanism that approach $100\%$. The high efficiency parameters are a direct result of the high energy contained in the flares and the extreme duration of the dipole component, which forces either slow spin periods or low magnetic fields. We find that even with the inclusion of significant fallback accretion, in all but a few cases it is energetically challenging to produce prompt emission, afterglow and giant flares within the constraints of the rotational energy budget of a magnetar.
\end{abstract}

\section{Introduction}
\label{sec:intro}

Gamma-ray bursts (GRBs) are intense explosions that outshine any other source in the gamma-ray sky while they are active \citep{meszaros06}. They occur randomly throughout the Universe and are generally categorised into two types \citep{kouveliotou93}: short GRBs (SGRBs; lasting $<2$ seconds) and long GRBs (LGRBs; lasting $>2$ seconds)\footnote{\label{footnote:timing}Though the duration of the gamma-rays alone is not able to unambiguously distinguish between the two types, see \cite{bromberg13}.}. The launch of the \emph{Swift} satellite in 2004 \citep{gehrels04} facilitated a break-through in our understanding of GRB physics thanks to its rapid slewing capabilities allowing early and well-sampled observations of afterglows by the X-ray Telescope (XRT) \citep{burrows05a}. This led to the formation of a `canonical' X-ray afterglow model consisting of the following phases \citep{nousek06,obrien06}: (\emph{i}) a steep, early decay; (\emph{ii}) a plateau; (\emph{iii}) a late decay; (\emph{iv}) achromatic jet breaks; and (\emph{v}) flares. Phases (\emph{iv}) and (\emph{v}) do not always have to be present and flares are often superposed onto the plateau phase \citep{curran08}.

Flares are a dramatic rebrightening in the X-ray light curve that are seen $\sim30-10^{5}$ seconds after the burst trigger \citep{burrows05b,beniamini16} and are observed in approximately half of all GRBs detected by \emph{Swift}-XRT \citep{obrien06,curran08,swenson14}. \citet{margutti11} note that, observationally, there appears to be two different evolutions of X-ray flare luminosity with time. The average luminosity of flares occurring before $t=1000~{\rm s}$ decreases as $t^{-2.7}$, while the luminosity at later times decreases as $\sim t^{-1}$. Flares are characterised by a fast rise, exponential decay (FRED) profile. The fluence of the largest flares (so-called giant flares) is often comparable to the prompt emission, potentially indicating a common origin between the two \citep{chincarini10}. The presence of an underlying continuum that is unaffected by the flare (i.e.~the superposition of the flare on the plateau phase) indicates that the flares do not share an emission site with the afterglow \citep{chincarini10}, which is believed to be produced by the deceleration of forward shocks in the ambient medium. GRB 050502B contains the first and the largest flare to be observed, rebrightening by a factor of $\sim500$ above the continuum \citep{falcone06}. The additional energy release observed in giant flares like the one seen in GRB 050502B  provide a unique test to constrain the energy budget of GRBs.

There are a variety of models which have been suggested to explain the origin of flares, including: `patchy' shells \citep{meszaros98,kumar00b}; refreshed shocks \citep{rees98,zhang02}; and density fluctuations \citep{wang00,dai02}. The continued central engine (CE) activity model \citep{dai98,zhang02} is often favoured, since the characteristics of flares are similar to the prompt emission \citep{chincarini10}. The new-born millisecond magnetar is a concept that is competing with black holes as the source of power in GRBs, mainly due to its potential as a naturally long-lived central engine (see \citealt{bernardini15} for a review). In the magnetar model, the rotational energy of a highly-magnetised neutron star is tapped via interactions between its intense dipole field and the circumstellar environment \citep[see][]{zhang01}. This model has been successfully applied to short \citep{metzger08,gompertz13,rowlinson13} and long GRBs \citep{lyons10}. However, it has a strict energy upper limit imposed by the rotational energy reservoir of the neutron star. This is typically assumed to be $\approx 3\times10^{52}$~ergs for a $1.4$~M$_{\odot}$ neutron star with a $1$~ms spin period. The magnetar is expected to be spun down very rapidly during the prompt emission phase, thereby decreasing the amount of energy available to power a flare \citep{beniamini16}. However, fallback accretion may augment the magnetar energy budget, as it provides a mechanism to spin the magnetar back up. Recent work by \citet{beniamini17a} and \citet{metzger18} suggests that the extractable energy from an isolated magnetar usable in a GRB is even further reduced. They predict for the same neutron star, the limit would be $\sim2\times10^{51}~{\rm erg}$ making the need for fallback accretion even more severe.

In this paper, we investigate whether flares can be powered by the delayed on-set of a propeller regime \citep{piro11,gompertz14}, in which in-falling material is accelerated to super-Keplerian velocities via magneto-centrifugal slinging and is ejected from the system. A magnetic propeller provides a path to a smoother emission profile than can be achieved by direct accretion onto a compact object, matching the phenomenology of giant flares more closely. We maximise the available energy reservoir by feeding the disc with fallback accretion (which was successfully used to match the light curves of SGRBs with extended emission in \citealt{gibson17}), following models such as \citet{eksi05,rosswog07,kumar08,cannizzo11,parfrey16}.

In Section \ref{sec:method}, we briefly review the magnetar propeller with fallback accretion model used in \citet{gibson17}. We introduce our sample of long GRBs with significant X-ray flares in Section \ref{sec:sample} and present our results and discussion of the fitting procedure in Section \ref{sec:results}. We summarise our conclusions in Section \ref{sec:concs}.

\section{The Magnetar Model}
\label{sec:method}

The propeller regime is defined according to the relationship between the Alfv\'en radius (where the dynamics of the disc are strongly influenced by the magnetic field, $r_{\rm m}$) and the co-rotation radius (where matter in the disc orbits at the same rate as the rotation of the stellar surface, $r_{\rm c}$). When $r_{\rm c}>r_{\rm m}$, the disc is rotating faster than the magnetic field (assuming the field lines rotate rigidly with the stellar surface) and magnetic torques slow the in-falling material allowing it to accrete. The magnetar gains angular momentum and is spun-up causing $r_{\rm c}$ to migrate towards the magnetar. This also increases the rotation of the magnetic field lines, causing $r_{\rm m}$ to migrate outwards. This leads to the opposite case of $r_{\rm m}>r_{\rm c}$, so that the magnetic field is rotating faster than the disc. Material is therefore accelerated to Super-Keplerian velocities, via direct interaction with the neutron stars magnetic field, and propelled to the light cylinder radius before being ejected from the system \citep{piro11}. The magnetar loses angular momentum to the ejected material and is spun-down. This is the propeller regime.

In \citet{gibson17}, we expanded this basic model \citep{piro11,gompertz14} to include fallback accretion. This was used to successfully reproduce both the prompt emission energy and extended emission in our sample of SGRBs. Full details of the model and results can be found within \citet{gibson17}. We parameterised the fallback timescale as a fraction, $\epsilon$, of the viscous timescale of the disc such that $t_{\rm fb} = \epsilon t_{\nu}$. Similarly, the fallback mass budget was defined as a fraction, $\delta$, of the initial disc mass such that $M_{\rm fb} = \frac{3}{2}\delta M_{\rm D,i}$. A mass flow rate of material through the disc - accounting for accretion on to the magnetar, propellering out of the system, and fallback into the disc - was defined as follows:
\begin{equation}
\label{eq:mdotdisc}
\dot{M}_{\rm D}(t)=\dot{M}_{\rm fb}-\dot{M}_{\rm acc}-\dot{M}_{\rm prop},
\end{equation}
where the fallback rate is defined as:
\begin{equation}
\label{eq:mdotfb}
\dot{M}_{\rm fb}(t)=\frac{M_{\rm fb}}{t_{\rm fb}}{\left(\frac{t+t_{\rm fb}}{t_{\rm fb}}\right)}^{-\frac{5}{3}},\end{equation}
using the ballistic timescale of $t^{-5/3}$ from \cite{rosswog07}.

Equation (\ref{eq:mdotdisc}) and the angular frequency of the magnetar, $\omega(t)$, have been solved over time and from these values the propellered and dipole components of the luminosity (and hence the total luminosity) can be calculated as follows:
\begin{equation}
\label{eq:Ldip}
L_{\rm dip} = -\tau_{\rm dip}\omega
\end{equation}
\begin{equation}
\label{eq:Lprop}
L_{\rm prop}=-\tau_{\rm acc}\omega
\end{equation}
\begin{equation}
\label{eq:Ltot}
L_{\rm tot}=\frac{1}{f_{\rm B}}\left(\eta_{\rm prop}L_{\rm prop}+\eta_{\rm dip}L_{\rm dip}\right)
\end{equation}
where $\tau_{\rm acc}$ and $\tau_{\rm dip}$ are the accretion and dipole torques, respectively, which have been defined to be positive when the magnetar is spinning up and negative when it's spinning down. During  spin up phases,  $L_{\rm prop} $ is set to zero. The quantities $\eta_{\rm prop}$ and $\eta_{\rm dip}$ represent the efficiencies of the propeller and dipole emission components respectively and $1/f_{\rm B}$ is the beaming fraction.

We use a Markov chain Monte Carlo (MCMC) simulation package \citep{emcee} to find the optimal values for our $9$ free parameters: $B$ - magnetic field strength of the magnetar; $P_{\rm i}$ - spin period of the magnetar; $M_{\rm D,i}$ - disc mass; - $R_{\rm D}$ - disc radius; $\epsilon$ - fallback timescale fraction; $\delta$ - fallback mass budget fraction; $\eta_{\rm dip}$ - dipole energy to luminosity conversion efficiency; $\eta_{\rm prop}$ - propeller energy to luminosity conversion efficiency; and $1/f_{\rm B}$ - beaming fraction (please see Appendix \ref{sec:appendix} for a discussion of the correlations between these fitting parameters and why a degeneracy treatment is not required). These parameters are defined after the prompt phase has ceased, which has been arbitrarily chosen to be $t=1~{\rm s}$. We used $200$ `walkers' taking $50,000$ steps each and constructed a posterior probability distribution from a Gaussian log-likelihood function and a flat prior function \citep[using the parameter limits given in Table 4 in][]{gibson17}. Fixed parameters are the viscosity prescription, $\alpha=0.1$; the speed of sound in the accretion disc, $c_{\rm s}=10^{7}~{\rm cm~s^{-1}}$; the ratio $r_{\rm m}/r_{\rm lc}=0.9$, which prevents ejected material from exceeding the speed of light; and the dimensionless parameter $n=1$, which controls how rapidly the propeller emission becomes dominant.

\section{Sample of \emph{Swift} LGRBs with giant X-ray flares}
\label{sec:sample}

We have chosen a sample of $19$ LGRBs that exhibit significant flares in their X-ray afterglows to study. Since there is no consistent definition of a giant X-ray flare, we selected which LGRBs to study based on the sample rate of data through the duration of the flare. We require good data coverage near the peak of the flare and a reasonable constraint on the amplitude of the flare so that our fitting routine can properly constrain the free parameters, as such a prominent feature will drive the morphology of the fit.

The data were collected by \emph{Swift}-XRT \citep{gehrels04,burrows05a} and were processed by the UK \emph{Swift} Science Data Centre (UKSSDC\footnote{\label{footnote:swift}www.swift.ac.uk}; \citealt{evans07,evans09}). In order to produce bolometric, rest-frame light curves, the data underwent a cosmological $k$-correction \citep{bloom01} and were corrected for absorption using values in Table \ref{tab:kcorr}. For those GRBs with no observed redshift, the mean of the sample in \citet{salvaterra12} was used (i.e.~$z = 1.84$).

\begin{table}
\centering
\caption{The parameters required to perform a cosmological $k$-correction as described by \citet{bloom01}. $\Gamma$ is the photon index; $\sigma$ is the absorption coefficient calculated from the ratio of counts-to-flux (unabsorbed) to counts-to-flux (absorbed); and $z$ is the redshift given in the literature. For those GRBs with no observed redshift (marked with an $^{\rm *}$), the mean of the sample in \citet{salvaterra12} was used. $^{\rm a}$\citet{afonso11}; $^{\rm b}$\citet{mirabal06}; $^{\rm c}$\citet{berger06}; $^{\rm d}$\citet{fugazza06}; $^{\rm e}$\citet{bloom06}; $^{\rm f}$\citet{penacchioni13}; $^{\rm g}$\citet{cabrera11}; $^{\rm h}$\citet{elliott14}; $^{\rm i}$\citet{tanvir16}.}
\label{tab:kcorr}
\begin{tabular}{llll}
\hline
GRB & $\Gamma$ & $\sigma$ & $z$ \\
\hline
050502B & $1.907^{+0.125}_{-0.098}$ & $1.11$ & $5.2^{\rm a}$ \\ [2pt]
060124 & $1.91^{+0.06}_{-0.05}$ & $1.28$ & $2.297^{\rm b}$ \\ [2pt]
060526 & $1.98^{+0.17}_{-0.12}$ & $1.15$ & $3.21^{\rm c}$ \\ [2pt]
060904B & $2.05^{+0.15}_{-0.15}$ & $1.49$ & $0.703^{\rm d}$ \\ [2pt]
060929 & $3.5^{+1.0}_{-1.4}$ & $5.79$ & $1.84^{\rm *}$ \\ [2pt]
061121 & $1.82^{+0.06}_{-0.06}$ & $1.23$ & $1.314^{\rm e}$ \\ [2pt]
070520B & $2.5^{+0.8}_{-0.6}$ & $1.70$ & $1.84^{\rm *}$ \\ [2pt]
070704 & $2.3^{+0.5}_{-0.4}$ & $3.15$ & $1.84^{\rm *}$ \\ [2pt]
090621A & $2.09^{+0.26}_{-0.25}$ & $2.42$ & $1.84^{\rm *}$ \\ [2pt]
100619A & $2.30^{+0.16}_{-0.15}$ & $2.19$ & $1.84^{\rm *}$ \\ [2pt]
110709B & $2.01^{+0.06}_{-0.06}$ & $1.38$ & $0.75^{\rm f}$ \\ [2pt]
110801A & $1.99^{+0.11}_{-0.10}$ & $1.25$ & $1.858^{\rm g}$ \\ [2pt]
110820A & $2.5^{+0.6}_{-0.5}$ & $2.62$ & $1.84^{\rm *}$ \\ [2pt]
121123A & $1.85^{+0.11}_{-0.11}$ & $1.17$ & $1.84^{\rm *}$ \\ [2pt]
121217A & $1.97^{+0.11}_{-0.11}$ & $1.66$ & $3.1^{\rm h}$ \\ [2pt]
140817A & $1.803^{+0.103}_{-0.100}$ & $1.30$ & $1.84^{\rm *}$ \\ [2pt]
141031A & $1.85^{+0.32}_{-0.16}$ & $1.31$ & $1.84^{\rm *}$ \\ [2pt]
141130A & $2.0^{+0.4}_{-0.3}$ & $1.15$ & $1.84^{\rm *}$ \\ [2pt]
160425A & $2.47^{+0.20}_{-0.19}$ & $2.19$ & $0.555^{\rm i}$ \\
\hline
\end{tabular}
\end{table}

\section{Results and Discussion}
\label{sec:results}

\begin{figure*}
\includegraphics[width=\textwidth]{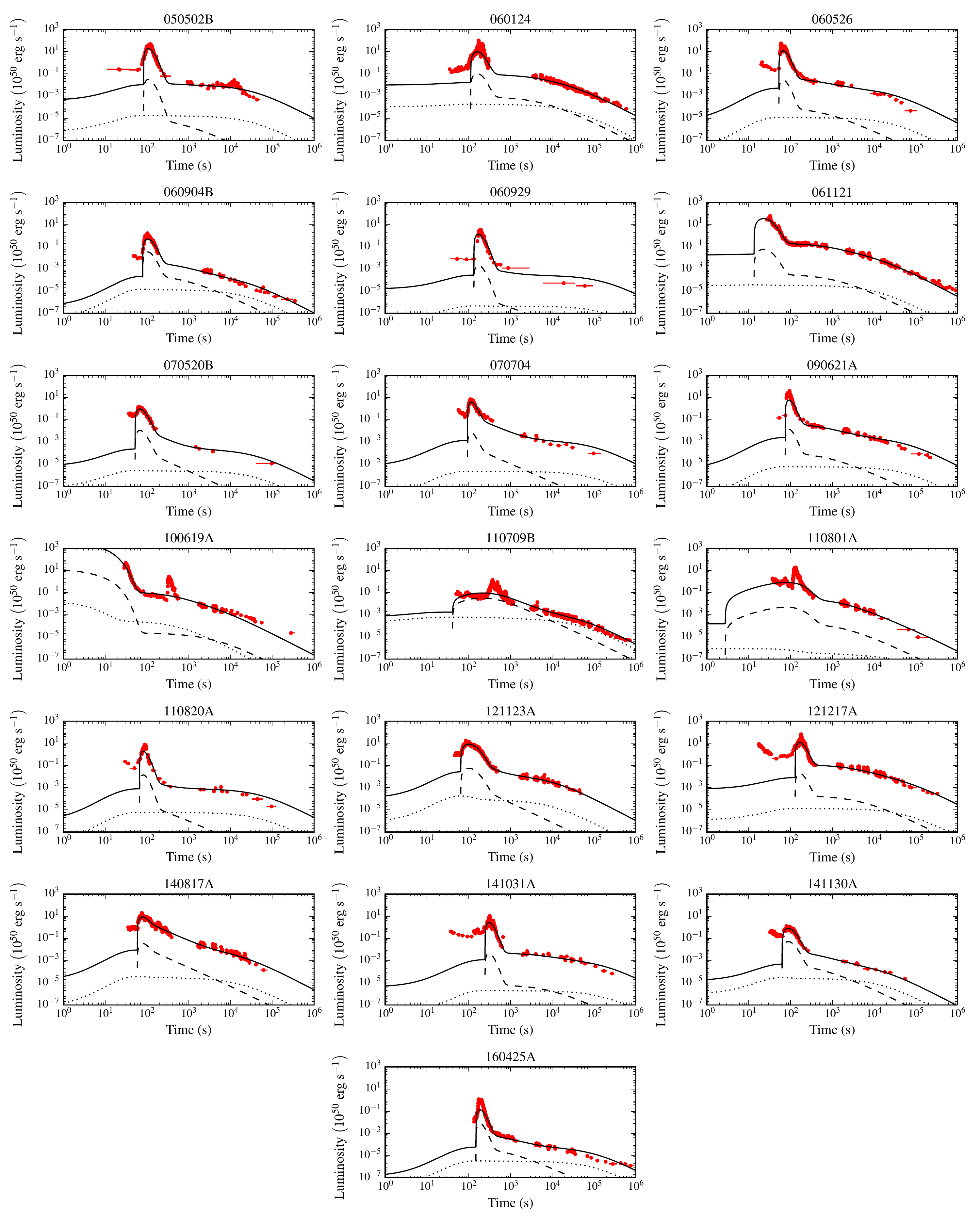}
\caption{Fits of magnetar propeller model with fallback accretion to LGRB with X-ray flare sample. Red points are \emph{Swift}-XRT data; solid, black line - total luminosity; dashed, black line - propeller luminosity; dotted, black line - dipole luminosity.}
\label{fig:results}
\end{figure*}

\begin{table*}
\centering
\caption{Parameters derived from fits shown in Fig.~\ref{fig:results} and the $\chi^{2}_{\rm red}$ goodness of fit statistic. Values marked with an [L] are a parameter limit.}
\begin{tabular}{lcccccccccc}
\hline
GRB & $B$ & $P_{\rm i}$ & $M_{\rm D,i}$ & $R_{\rm D}$ & $\epsilon$ & $\delta$ & $\eta_{\rm dip}$ & $\eta_{\rm prop}$ & $1/f_{\rm B}$ & $\chi^{2}_{\rm red}$ \\
& $\left(\times10^{15}~{\rm G} \right)$ & (ms) & $\left(\times10^{-2}~{\rm M}_{\odot}\right)$ & (km) & & & (\%) & (\%) & \\
\hline
050502B & $0.47^{+0.02}_{-0.02}$ & $4.00^{+0.27}_{-0.22}$ & $9.99^{+0.01}_{-0.03}$ & $217^{+1}_{-1}$ & $23.24^{+250.16}_{-23.12}$ & $\left(2.14^{+5.25}_{-1.10}\right)\times10^{-5}$ & $1$[L] & $100$[L] & $599^{+1}_{-3}$ & $8$ \\ [2pt]
060124 & $0.165^{+0.006}_{-0.003}$ & $0.70^{+0.03}_{-0.01}$ & $9.98^{+0.02}_{-0.10}$ & $417^{+4}_{-5}$ & $118.79^{+15.78}_{-13.93}$ & $\left(9.83^{+0.87}_{-0.87}\right)\times10^{-3}$ & $1$[L] & $99^{+1}_{-4}$ & $91^{+7}_{-4}$ & $21$ \\ [2pt]
060526 & $0.66^{+0.01}_{-0.01}$ & $9.97^{+0.03}_{-0.13}$ & $9.98^{+0.02}_{-0.09}$ & $120^{+1}_{-1}$ & $45.92^{+15.89}_{-13.34}$ & $\left(2.32^{+0.42}_{-0.38}\right)\times10^{-4}$ & $1$[L] & $100$[L] & $430^{+17}_{-16}$ & $20$ \\ [2pt]
060904B & $0.74^{+0.01}_{-0.01}$ & $9.94^{+0.06}_{-0.25}$ & $9.97^{+0.03}_{-0.14}$ & $225^{+2}_{-2}$ & $30.66^{+8.94}_{-6.91}$ & $\left(9.26^{+1.60}_{-1.43}\right)\times10^{-4}$ & $1$[L] & $100$[L] & $14^{+1}_{-1}$ & $10$ \\ [2pt]
060929 & $0.551^{+0.002}_{-0.002}$ & $10$[L] & $2.58^{+0.02}_{-0.02}$ & $329^{+1}_{-1}$ & $10.20^{+3.05}_{-2.72}$ & $\left(2.58^{+0.62}_{-0.63}\right)\times10^{-5}$ & $1$[L] & $100$[L] & $599^{+1}_{-3}$ & $419$ \\ [2pt]
061121 & $1.15^{+0.29}_{-0.22}$ & $2.82^{+0.51}_{-0.39}$ & $0.96^{+0.19}_{-0.15}$ & $84^{+2}_{-2}$ & $127.09^{+16.68}_{-17.55}$ & $\left(1.74^{+0.34}_{-0.33}\right)\times10^{-2}$ & $2^{+1}_{-1}$ & $98^{+2}_{-8}$ & $588^{+12}_{-48}$ & $3$ \\ [2pt]
070520B & $1.01^{+0.07}_{-0.05}$ & $9.79^{+0.20}_{-0.84}$ & $3.31^{+0.90}_{-0.74}$ & $142^{+4}_{-3}$ & $0.34^{+2.53}_{-0.24}$ & $\left(1.43^{+2.06}_{-1.06}\right)\times10^{-3}$ & $1$[L] & $99^{+1}_{-4}$ & $86^{+29}_{-23}$ & $20$ \\ [2pt]
070704 & $0.43^{+0.01}_{-0.01}$ & $9.99^{+0.01}_{-0.03}$ & $7.24^{+0.12}_{-0.12}$ & $188^{+1}_{-1}$ & $2.02^{+0.93}_{-0.87}$ & $\left(6.57^{+2.43}_{-1.28}\right)\times10^{-4}$ & $1$[L] & $100$[L] & $599^{+1}_{-5}$ & $62$ \\ [2pt]
090621A & $0.458^{+0.004}_{-0.004}$ & $9.99^{+0.01}_{-0.04}$ & $9.97^{+0.03}_{-0.11}$ & $154^{+1}_{-1}$ & $47.24^{+5.00}_{-4.70}$ & $\left(3.54^{+0.22}_{-0.21}\right)\times10^{-4}$ & $1$[L] & $100$[L] & $423^{+12}_{-11}$ & $91$ \\ [2pt]
100619A & $7.19^{+2.26}_{-1.72}$ & $1.53^{+0.19}_{-0.19}$ & $7.04^{+2.79}_{-2.34}$ & $50$[L] & $996.17^{+3.70}_{-16.08}$ & $\left(1.10^{+0.14}_{-0.13}\right)\times10^{-3}$ & $1$[L] & $83^{+16}_{-24}$ & $394^{+194}_{-171}$ & $124$ \\ [2pt]
110709B & $0.13^{+0.47}_{-0.01}$ & $0.69$[L] & $9.93^{+0.07}_{-9.82}$ & $51^{+1798}_{-1}$ & $30.93^{+108.34}_{-30.82}$ & $0.28^{+12.57}_{-0.04}$ & $4^{+4}_{-3}$ & $37^{+42}_{-21}$ & $3^{+65}_{-2}$ & $44$ \\ [2pt]
110801A & $2.05^{+0.27}_{-0.51}$ & $8.47^{+1.47}_{-3.55}$ & $0.44^{+0.47}_{-0.13}$ & $737^{+17}_{-16}$ & $23.29^{+3.98}_{-3.38}$ & $\left(7.28^{+0.77}_{-0.57}\right)\times10^{-2}$ & $1$[L] & $91^{+9}_{-27}$ & $174^{+103}_{-112}$ & $37$ \\ [2pt]
110820A & $0.48^{+0.01}_{-0.01}$ & $9.99^{+0.01}_{-0.06}$ & $9.75^{+0.23}_{-0.59}$ & $134^{+1}_{-1}$ & $6.13^{+9.64}_{-5.68}$ & $\left(2.66^{+6.49}_{-0.94}\right)\times10^{-5}$ & $1$[L] & $100$[L] & $134^{+14}_{-7}$ & $137$ \\ [2pt]
121123A & $1.57^{+0.03}_{-0.04}$ & $9.77^{+0.22}_{-1.00}$ & $9.78^{+0.22}_{-0.79}$ & $343^{+4}_{-5}$ & $0.18^{+0.59}_{-0.08}$ & $\left(1.05^{+0.57}_{-0.63}\right)\times10^{-2}$ & $5^{+3}_{-3}$ & $62^{+36}_{-36}$ & $152^{+212}_{-57}$ & $5$ \\ [2pt]
121217A & $0.29^{+0.01}_{-0.01}$ & $2.78^{+0.16}_{-0.14}$ & $9.99^{+0.01}_{-0.04}$ & $324^{+1}_{-1}$ & $127.61^{+9.78}_{-9.10}$ & $\left(2.76^{+0.15}_{-0.14}\right)\times10^{-3}$ & $1$[L] & $100$[L] & $598^{+2}_{-9}$ & $27$ \\ [2pt]
140817A & $0.85^{+0.03}_{-0.04}$ & $9.46^{+0.52}_{-1.56}$ & $9.94^{+0.06}_{-0.26}$ & $164^{+4}_{-6}$ & $1.40^{+0.67}_{-1.28}$ & $\left(1.90^{+7.41}_{-0.32}\right)\times10^{-2}$ & $2^{+1}_{-1}$ & $72^{+26}_{-28}$ & $249^{+156}_{-70}$ & $9$ \\ [2pt]
141031A & $0.26^{+0.01}_{-0.01}$ & $9.43^{+0.55}_{-1.21}$ & $9.98^{+0.02}_{-0.10}$ & $524^{+3}_{-3}$ & $144.96^{+53.20}_{-42.31}$ & $\left(2.29^{+0.52}_{-0.48}\right)\times10^{-4}$ & $1$[L] & $100$[L] & $597^{+3}_{-11}$ & $21$ \\ [2pt]
141130A & $0.94^{+0.35}_{-0.39}$ & $5.35^{+4.46}_{-3.49}$ & $8.01^{+1.93}_{-5.80}$ & $228^{+7}_{-11}$ & $19.28^{+33.27}_{-15.60}$ & $\left(1.42^{+0.63}_{-0.70}\right)\times10^{-3}$ & $1$[L] & $89^{+11}_{-30}$ & $16^{+73}_{-9}$ & $10$ \\ [2pt]
160425A & $0.349^{+0.003}_{-0.004}$ & $9.99^{+0.01}_{-0.06}$ & $9.97^{+0.03}_{-0.11}$ & $319^{+1}_{-1}$ & $22.87^{+3.44}_{-3.10}$ & $\left(1.95^{+0.14}_{-0.14}\right)\times10^{-4}$ & $1$[L] & $100$[L] & $17.3^{+0.6}_{-0.4}$ & $113$ \\
\hline
\end{tabular}
\label{tab:results}
\end{table*}

The best fits of the magnetar propeller with fallback accretion model to our LGRB giant flare sample are presented in Fig.~\ref{fig:results}. The model provides a reasonable fit to the morphology of the data across the sample, recreating the height and shape of the flare and fitting the emission `tail' in $16$ out of $19$ GRBs. However in general terms, the model is struggling to meet the general energy budget of the sample which causes some of the parameters to be forced to the extremes of their allowed parameter space \citep[see Table 4 in][]{gibson17}. The model consistently misses the emission preceding the flare, falling 1-2 orders of magnitude lower than the data. However, this emission is most likely the tail end of the prompt spike, which we do not fit in this paper. The fits which performed the most poorly are to GRBs 100619A, 110709B and 110801A. In the case of GRB 100619A, the model has missed the second flare entirely in favour of fitting to the first flare\footnote{\label{footnote:100619A}GRB 100619A exhibits a double flare which is most obvious in the joint BAT and XRT light curve from the UKSSDC's Burst Analyser found here: http://www.swift.ac.uk/burst\_analyser/424998.}. Currently, our model is unable to fit multiple events like this as it does not contain an underlying flaring mechanism. Instead it describes a large release of energy that fits the general energetics of large flares. The closest approximation to multiple flares our model is currently capable of is a `stuttering' type burst (see \citealt{gompertz14} and \citealt{gibson17} for details of burst types). Double flares like this could be achieved using models such as `clumpy' accretion \citep[e.g.][]{dallosso17}, a self-criticality regime of magnetic reconnection \citep[similar to solar flares, e.g.][]{wang13}, or modulating the fallback rate to no longer be a smooth profile. As discussed in \citet{gibson17} in the context of the early time prompt emission, the model struggles to replicate short-timescale variability in GRBs 110709B and 110801A, instead `smoothing' through the main flare and the smaller, preceding flare. This is another feature that may be achievable with a `clumpy' accretion model, self-critical solar flare-like activity, or a modulated fallback rate. Mass would be delivered intermittently, causing outbursts as opposed to the smooth feeding currently modelled here.

The parameter values derived from the best-fitting models are presented in Table \ref{tab:results}. Across the sample, we have generally found low magnetic fields and slow initial spin periods, indicating that the propeller mechanism would not be that strong. A low $B$-field and fast spin period, or a high $B$-field and slow spin period have previously been shown to be necessary for an effective propeller \citep{rowlinson13,gompertz14,gibson17}. The driving factor behind these parameters is likely to be the duration of the dipole emission. The plateau duration is given by \citep[cf.][]{zhang01}:
\begin{equation}
T_{\rm em} = 10^{3} I_{45} B_{\rm p,15}^{-2} P_{\rm i,0}^2 R_{10}^{-6} \mbox{ s,}
\end{equation}
where $I_{45}$ is the moment of inertia of the neutron star in units of $10^{45}$~g~cm$^2$, $B_{\rm p,15}$ is the dipole field strength in units of $10^{15}$~G, $P_{\rm i,0}$ is the spin period in ms and $R_{10}^{-6}$ is the neutron star radius is units of $10$~km. The dipole emission in our sample typically lasts $\sim 10^5$~s, and assuming $I_{45} = R_{10}^{-6} = 1$, this requires $B_{\rm p,15}^{-2} P_{\rm i,0}^2 \approx 100$. From this we can clearly see that either $B$ must be low, $P$ must be high, or a combination of the two.

\begin{figure}
\includegraphics[width=\columnwidth]{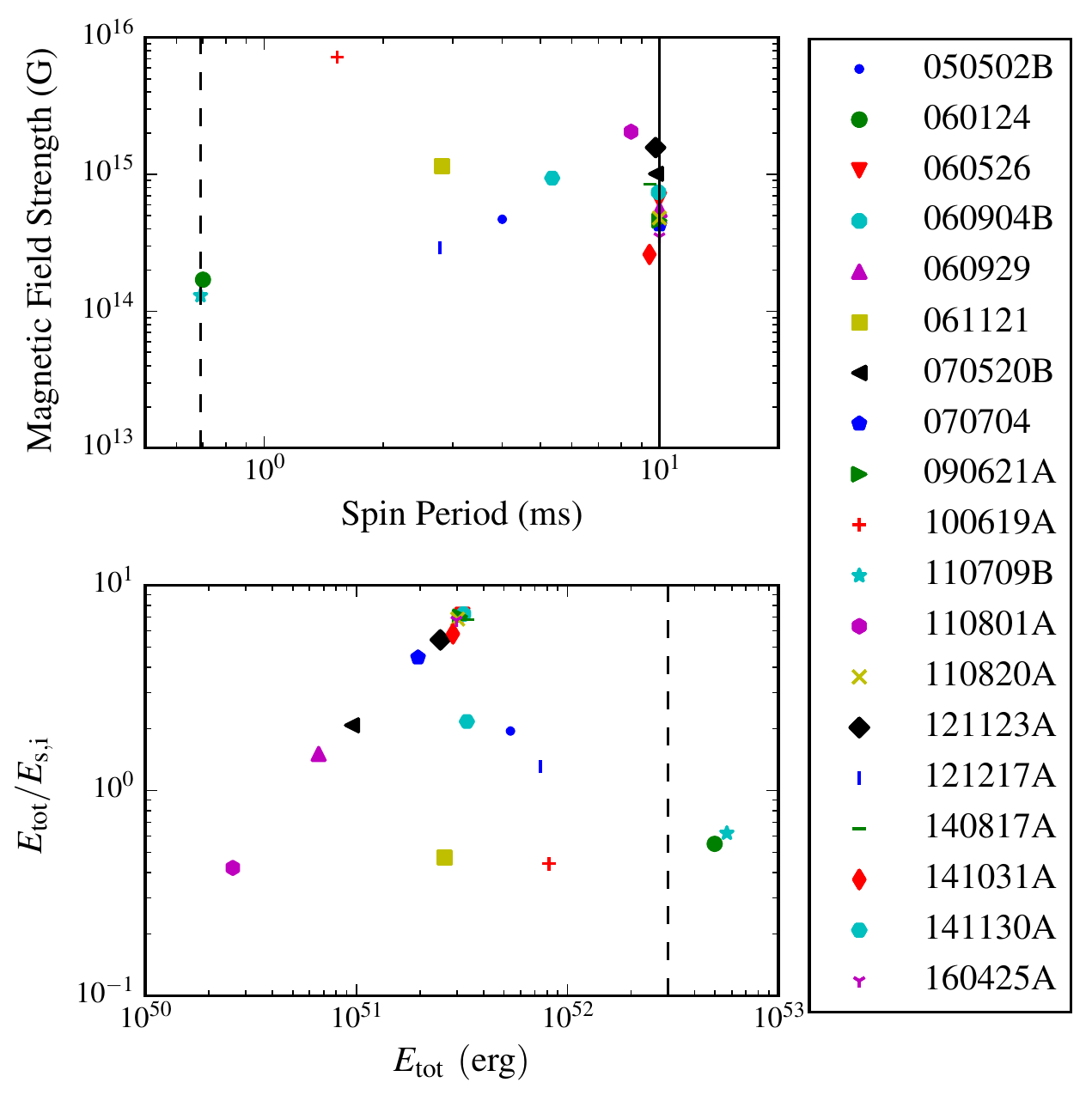}
\caption{\textbf{Top panel:} A plot of the magnetic field strength, $B$, against the initial spin period $P_{\rm i}$ of the LGRB giant flare sample. The solid line indicates the upper limit of $10~{\rm ms}$ and the dashed line indicates the lower, break-up limit of $0.69~{\rm ms}$ for a collapsar \citep{lattimer04}. Error bars were not included for clarity. \textbf{Bottom panel:} A plot showing the ratio of the total energy emitted to the initial spin energy, $E_{\rm tot}/E_{\rm s,i}$, against the total energy, $E_{\rm tot}$. The total energy emitted through radiation is calculated by integrating $L_{\rm dip}+L_{\rm prop}$ over time of each best fit model. The initial spin energy is given by $E_{\rm s,i}=\frac{1}{2}I\omega_{i}^{2}$, where $I=\frac{4}{5}MR^{2}$ is the moment of inertia of the magnetar and $\omega_{i}$ is the initial angular frequency. The dashed line represents the rotational energy reservoir ($\approx3\times10^{52}$ ergs) for a $1.4~{\rm M}_{\odot}$ neutron star with a $1$ ms spin period.}
\label{fig:energy}
\end{figure}

The top panel of Fig.~\ref{fig:energy} shows where the LGRB giant flare sample lies on the spin period-magnetic field plane. $11$ GRBs are clustered against the $10~{\rm ms}$ upper parameter limit and the majority of the sample have a magnetic field of the order of $1\times10^{15}~{\rm G}$ or less, which are consistent with the theoretical predictions for a magnetar \citep{giacomazzo13,mereghetti15,rea15}. The bottom panel shows where the sample lies in energy space as a fraction of the initial spin energy. The cluster of $9$ GRBs at the top of the plot all have low $B$-fields, $\lesssim 1\times10^{15}~{\rm G}$, and slow spin periods, $\simeq 10~{\rm ms}$,  one of the necessary conditions for an effective propeller.  The $2$ GRBs over the upper limit of the spinning neutron star model (GRBs 060124 and 110709B) both have low $B$ and $P_{\rm i}$ values and, therefore, have an ineffective propeller mechanism. Hence, the fallback has to compensate to provide the remaining energy. Table \ref{tab:results} shows us that both of these fits exhibit a significant fraction of the initial disc mass falling back on long timescales compared to the viscous timescale.

Since $E \propto P_{\rm i}^{-2}$, initial spin periods of $\sim 10$~ms reduce the total available energy by a factor of $100$. Conversely, in most cases the efficiency of the propeller $\eta_{\rm prop}$ is forced to $100$ per cent, likely in order to compensate for the low total available energy in the model. The mean beaming factor is $303$, translating into a jet opening angle of $\theta \approx 4.65^{\circ}$. This narrow beam is likely a further symptom of a system short of energy. We note at this point that alternative sources of energy which have been ignored here may also make a significant contribution and lower the energy requirements for our model. In particular, we neglect the contribution of the synchrotron emission from the afterglow as the forward shock driven by the initial explosion decelerates. Reprocessing of the dipole radiation in the forward shock will also allow longer-lived afterglow emission and enable lower values of $P$ and/or higher values of $B$ compared to the simplified treatment of the dipole applied here \citep[see e.g.][]{gompertz15}. Although the relatively long initial spin periods found in the fits are primarily resultant from the need to fit the dipole emission component, we note here that they would be broadly consistent with an episode of magnetar spin down during the prompt emission phase.

The wide range of values in the fallback parameters $\epsilon$ and $\delta$ spin the magnetar up at a later period, producing a more effective propeller mechanism. We also find more initial disc masses at the upper parameter limit and with smaller disc radii, which shows that the model is attempting to extract as much energy as possible through high accretion rates to fuel the flares. The dipole and propeller efficiencies, $\eta_{\rm dip}$ and $\eta_{\rm prop}$, are often pushed to their lowest and highest parameter limits respectively. This is because the flares produce such large flux increases above the smooth continuum that the model can only reproduce a rise and drop-off of this magnitude by having extremely different efficiencies for the dipole and propeller luminosities, despite this not being observationally consistent.

Although having a mechanism with $>50\%$ efficiency is likely unphysical and observationally inconsistent, it was found to be necessary for both efficiencies to be allowed to vary up to $100\%$ in order for the MCMC simulation to find an acceptable fit with constrained parameters. We ran the MCMC with different combinations of upper limits on the efficiencies and the $\chi^{2}_{\rm red}$ values of 2 arbitrarily chosen runs are presented in Table \ref{tab:chisq} along with the values for the fits in Fig.~\ref{fig:results} for comparison. In each case over all the runs, the MCMC was not able to constrain a value for $\eta_{\rm prop}$ since each value in the allowed limits had an equally poor $\chi^{2}_{\rm red}$ value as every other. In addition, we ran MCMC simulations that ignored the first $10$~s after trigger, which is typically unconstrained by data. We found consistently poor fits, indicating that our conclusions are not dominated by the early (unconstrained) part of the light curve. The dominant limiting factor appears to be the long duration of the emission demanding extreme values of $B$ and $P$, which forces the other parameters to work around them.

\begin{table}
\centering
\caption{The $\chi^{2}_{\rm red}$ values for fits with different upper limits on the dipole and propeller efficiencies ($\eta_{\rm dip}$ and $\eta_{\rm prop}$, respectively).}
\begin{tabular}{lccc}
\hline
 & & $\chi^{2}_{\rm red}$ & \\ [2pt]
 & $\eta_{\rm dip}=100\%$ & $\eta_{\rm dip}=100\%$ & $\eta_{\rm dip}=50\%$ \\
GRB & $\eta_{\rm prop}=100\%$ & $\eta_{\rm prop}=50\%$ & $\eta_{\rm prop}=50\%$ \\
\hline
050502B & 8 & 304 & 891 \\
060124 & 21 & 5,633 & 3,477 \\
060526 & 20 & 3,066 & 6,139 \\
060904B & 10 & 125,262 & 1,125 \\
060929 & 419 & 119,199 & 155,054 \\
061121 & 3 & 3,533 & 7,841 \\
070520B & 20 & 279,130 & 381,492 \\
070704 & 62 & 63,645 & 24,108 \\
090621A & 91 & 10,819 & 26,605 \\
100619A & 124 & 11,827 & 192,847 \\
110709B & 44 & 324,442 & 66,210 \\
110801A & 37 & 2,393 & 5,189 \\
110820A & 137 & 75,866 & 146,362 \\
121123A & 5 & 18,788 & 88,012 \\
121217A & 27 & 1,607 & 3,425 \\
140817A & 9 & 958 & 14,149 \\
141031A & 21 & 52,994 & 3,226 \\
141130A & 10 & 214,850 & 216,761 \\
160425A & 113 & 85,887 & 212,457 \\
\hline
\end{tabular}
\label{tab:chisq}
\end{table}

\begin{table}
\centering
\caption{Lorentz factor values of the X-ray flare sample, $\Gamma_{\rm X}$, calculated using the $\Gamma_{\gamma}-E_{\gamma{\rm ,iso}}$ relation in \citet{lu12} and \citet{mu16}. The first column corresponds to the flare Lorentz factor calculated from the \emph{Swift} data, while the second column corresponds to the flare Lorentz factor calculated from the best fitting models. Since the model misses the second, well-defined flare in GRB 100619A, we have not provided a value of $\Gamma_{\rm X,model}$ for it. GRBs marked with an $^*$ have no redshift and the mean of the sample in \citet{salvaterra12} was used.\label{tab:gamma}}
\begin{tabular}{lcc}
\hline
GRB & $\Gamma_{\rm X,data}$ & $\Gamma_{\rm X,model}$ \\
\hline
050502B & $145.75\pm0.56$ & $105.67$ \\
060124 & $187.85\pm0.59$ & $84.43$ \\
060526 & $153.04\pm0.52$ & $90.63$ \\
060904B & $46.42\pm0.16$ & $30.71$ \\
060929$^*$ & $59.06\pm0.11$ & $43.04$ \\
061121 & $158.54\pm0.50$ & $130.40$ \\
070520B$^*$ & $44.28\pm0.12$ & $38.04$ \\
070704$^*$ & $75.49\pm0.17$ & $61.58$ \\
090621A$^*$ & $135.58\pm0.37$ & $70.55$ \\
100619A$^*$ & $55.24\pm0.15$ & $-$ \\
110709B & $43.47\pm0.13$ & $16.99$ \\
110801A & $104.93\pm0.33$ & $36.61$ \\
110820A$^*$ & $77.57\pm0.19$ & $48.40$ \\
121123A$^*$ & $103.35\pm0.34$ & $80.52$ \\
121217A & $161.01\pm0.49$ & $90.90$ \\
140817A$^*$ & $107.60\pm0.35$ & $81.28$ \\
141031A$^*$ & $86.33\pm0.27$ & $54.81$ \\
141130A$^*$ & $42.54\pm0.14$ & $36.24$ \\
160425A & $41.62\pm0.10$ & $19.75$ \\
\hline
\end{tabular}
\end{table}

Table \ref{tab:gamma} presents the values of the Lorentz factors for the X-ray flares in our sample, $\Gamma_{\rm X}$. These have been calculated using Equation (\ref{eq:mu}), which comes from \citet{lu12} and \citet{mu16}.
\begin{equation}
\label{eq:mu}
\log \Gamma_{\rm X} = (2.27\pm0.04) + (0.34\pm0.03) \log L_{{\rm X,p,52}}
\end{equation}
where $L_{{\rm X,p,52}}$ is the peak luminosity of the flare in units of $10^{52}~{\rm erg}~{\rm s}^{-1}$.

We find our flare sample  calculated from the data, $\Gamma_{\rm X,data}$, is broadly consistent with the findings of \citet{peng14} where $\Gamma_{\rm X}$ takes values of around $60 \sim 150$. Whereas, the Lorentz factors calculated from the best fitting models, $\Gamma_{\rm X,model}$, range from $\sim16-130$. While this includes the the majority of the range indicated by Peng et al.~(2014), the values are often lower than those required by the data especially in the case of the most powerful flares, e.g. GRB 060124. Since the model cannot produce Lorentz factors much greater than $\sim 100$, this further highlights that it is struggling to reach the energies demanded of it by the data.

\section{Conclusions}
\label{sec:concs}
Due to their similarity to the prompt emission, giant X-ray flares in LGRBs are often considered to be evidence of continuing central engine activity. In this paper, we test the feasibility of one of the most natural long-lived central engines: the magnetar model, in which the rotational energy of a highly-magnetised millisecond neutron star is released to the surrounding environment via its intense dipole field. Our model for flaring is a magnetic propeller, which accelerates local material via magneto-centrifugal slinging and ejects it from the system. The magnetar is fed by fallback accretion, which maximises the available energy. We provide fits to a sample of $19$ LGRBs with giant flares in their X-ray light curves using MCMC simulations.

Our results show that despite a good phenomenological match of the model to the data, in all but a few cases it is very energetically challenging to explain giant flares in LGRBs using a magnetar alone, especially given the further reduction of usable extracted energy predicted by \citet{beniamini17a} and \citet{metzger18}. This has strong implications for any models trying to explain LGRB prompt emission or late X-ray plateaux \citep{beniamini17b} with a magnetar, as the rotational energy budget appears to not be sufficient for flares without extra emission components or substantial fallback. However, the energy constraints may be lessened somewhat by the inclusion of the standard synchrotron afterglow and the reprocessing of the dipole emission in the forward shock.

\section*{Acknowledgements}

SLG would like to acknowledge funding from the Weizmann Institute and the University of Leicester. BPG has received funding from the European Research Council (ERC) under the European Union's Horizon 2020 research and innovation programme (grant agreement no 725246, TEDE, PI A. Levan). PTO would like to acknowledge funding from STFC. This research used the ALICE High Performance Computing Facility at the University of Leicester. The work makes use of data supplied by the UKSSDC at the University of Leicester and the \emph{Swift} satellite. \emph{Swift}, launched in November 2004, is a NASA mission in partnership with the Italian Space Agency and the UK Space Agency. \emph{Swift} is managed by NASA Goddard. Penn State University controls science and flight operations from the Mission Operations Centre in University Park, Pennsylvania. Los Alamos National Laboratory provides gamma-ray imaging analysis.


\bibliographystyle{mnras}
\bibliography{mylibrary}


\appendix

\section{Correlations between Fitting Parameters}
\label{sec:appendix}

In this Appendix, we address why we have used $9$ fitting parameters without any contingency for degeneracies between them.

\begin{figure*}
\centering
\includegraphics[width=\textwidth]{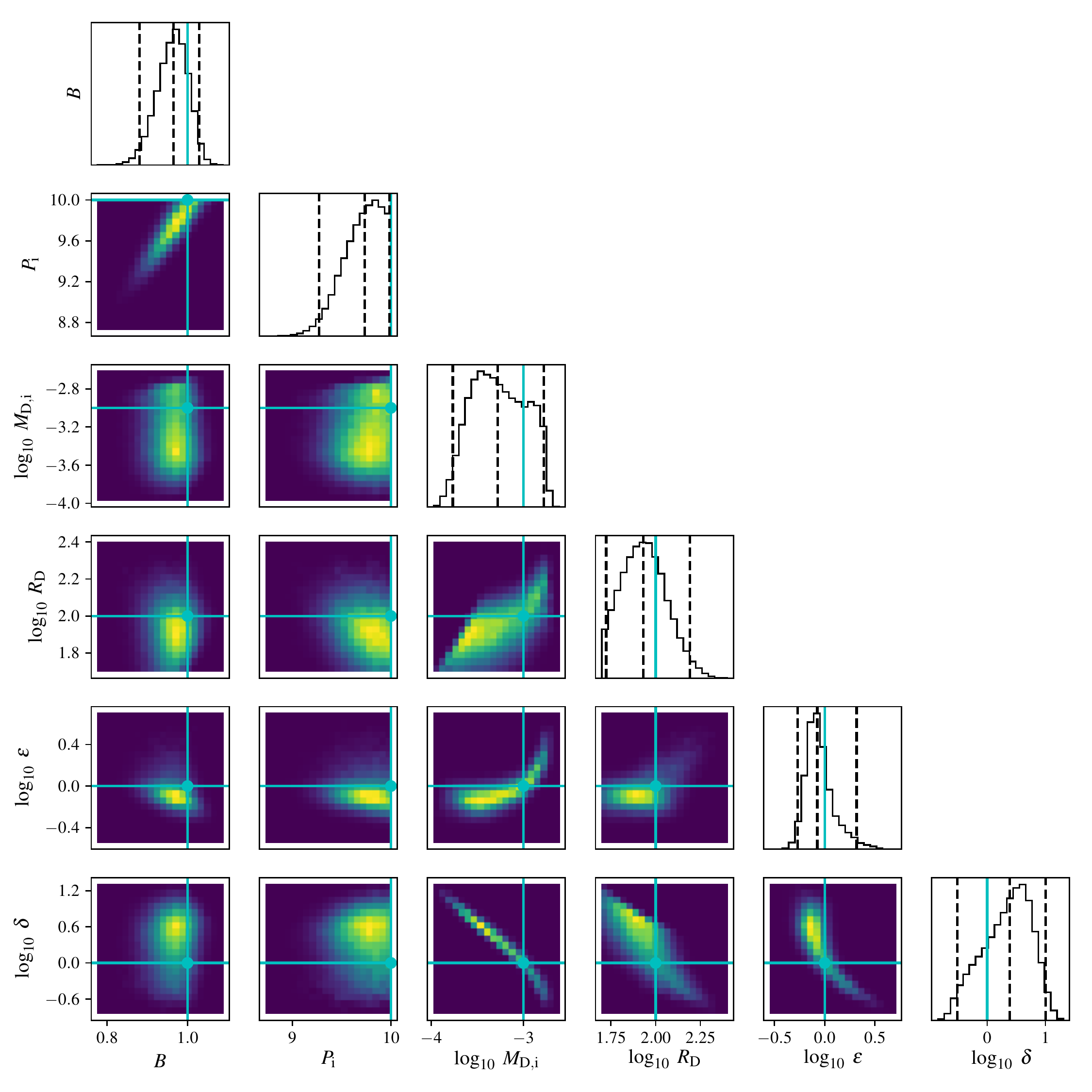}
\caption{2-D histograms showing the correlations between fitting parameters for a fit to a `humped' synthetic burst. The blue lines and points indicate the true values of the synthetic burst. The 1-D histograms show the sampled posterior distribution for each parameter. The dashed lines indicate the median an $\pm2\sigma$ values of the sampled posterior distribution and the blue line indicates the true value of the synthetic curve.}
\label{fig:humped}
\end{figure*}

\begin{figure*}
\centering
\includegraphics[width=\textwidth]{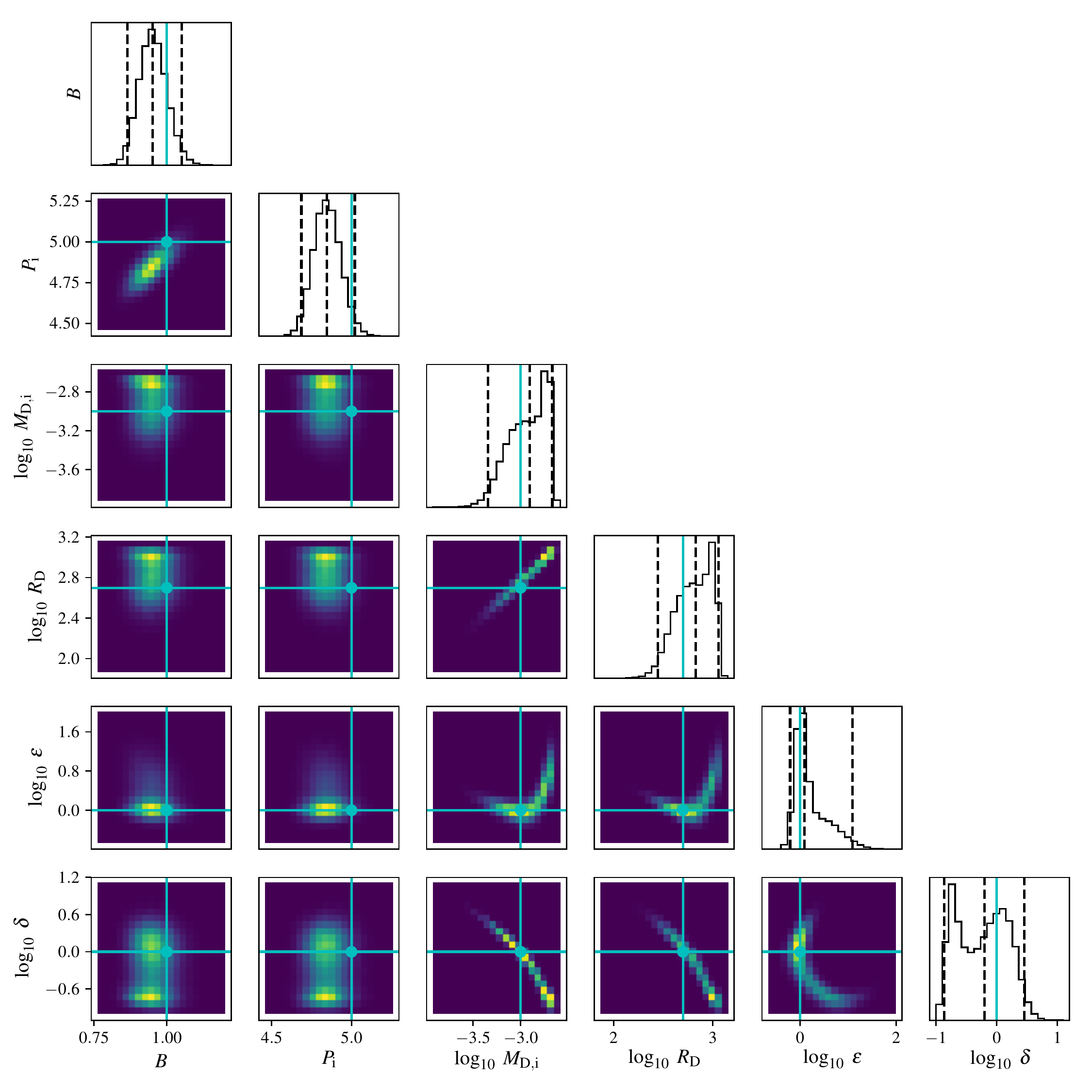}
\caption{2-D histograms showing the correlations between fitting parameters for a fit to a `classic' synthetic burst. The blue lines and points indicate the true values of the synthetic burst. The 1-D histograms show the sampled posterior distribution for each parameter. The dashed lines indicate the median an $\pm2\sigma$ values of the sampled posterior distribution and the blue line indicates the true value of the synthetic curve.}
\label{fig:classic}
\end{figure*}

\begin{figure*}
\centering
\includegraphics[width=\textwidth]{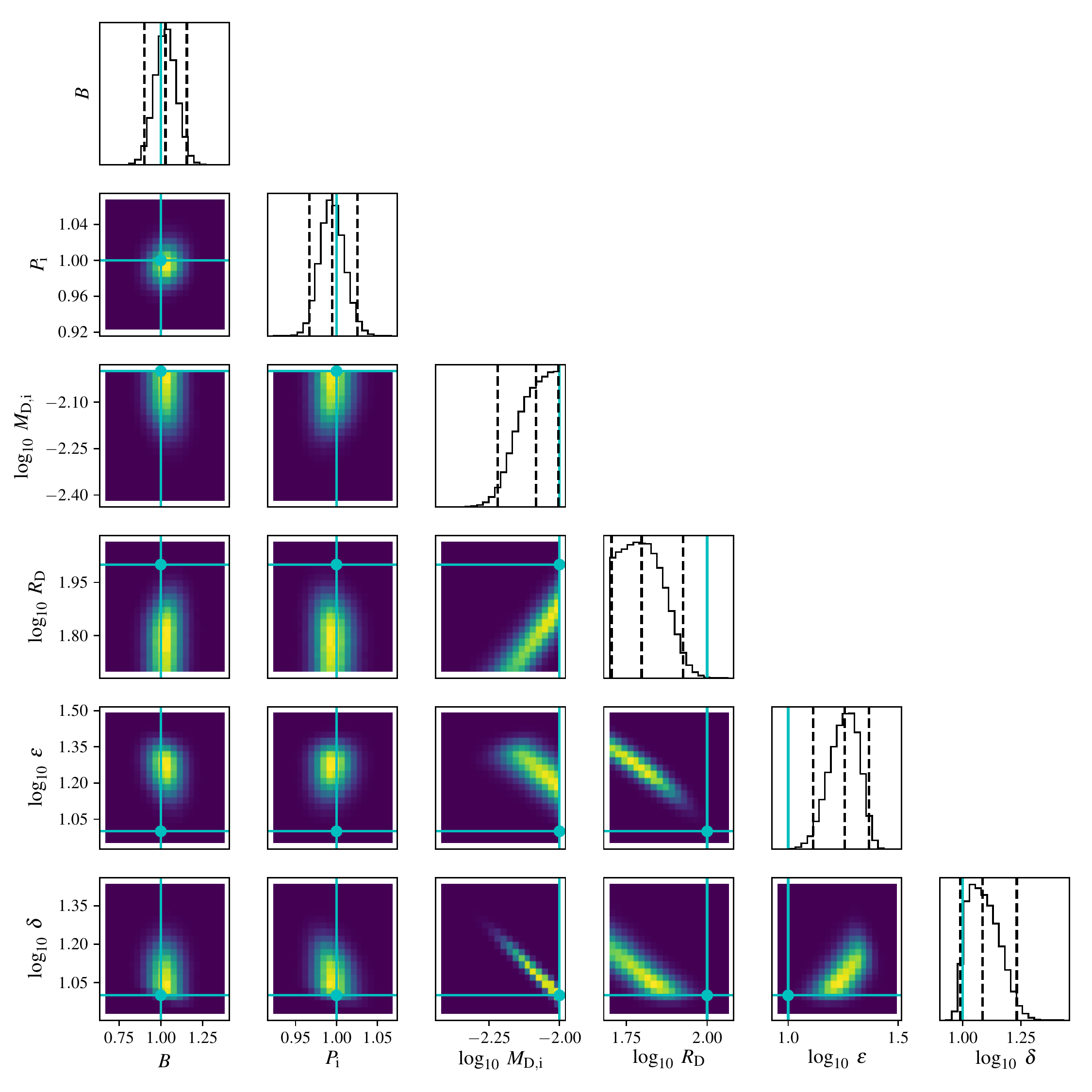}
\caption{2-D histograms showing the correlations between fitting parameters for a fit to a `sloped' synthetic burst. The blue lines and points indicate the true values of the synthetic burst.The 1-D histograms show the sampled posterior distribution for each parameter. The dashed lines indicate the median an $\pm2\sigma$ values of the sampled posterior distribution and the blue line indicates the true value of the synthetic curve.}
\label{fig:sloped}
\end{figure*}

\begin{figure*}
\centering
\includegraphics[width=\textwidth]{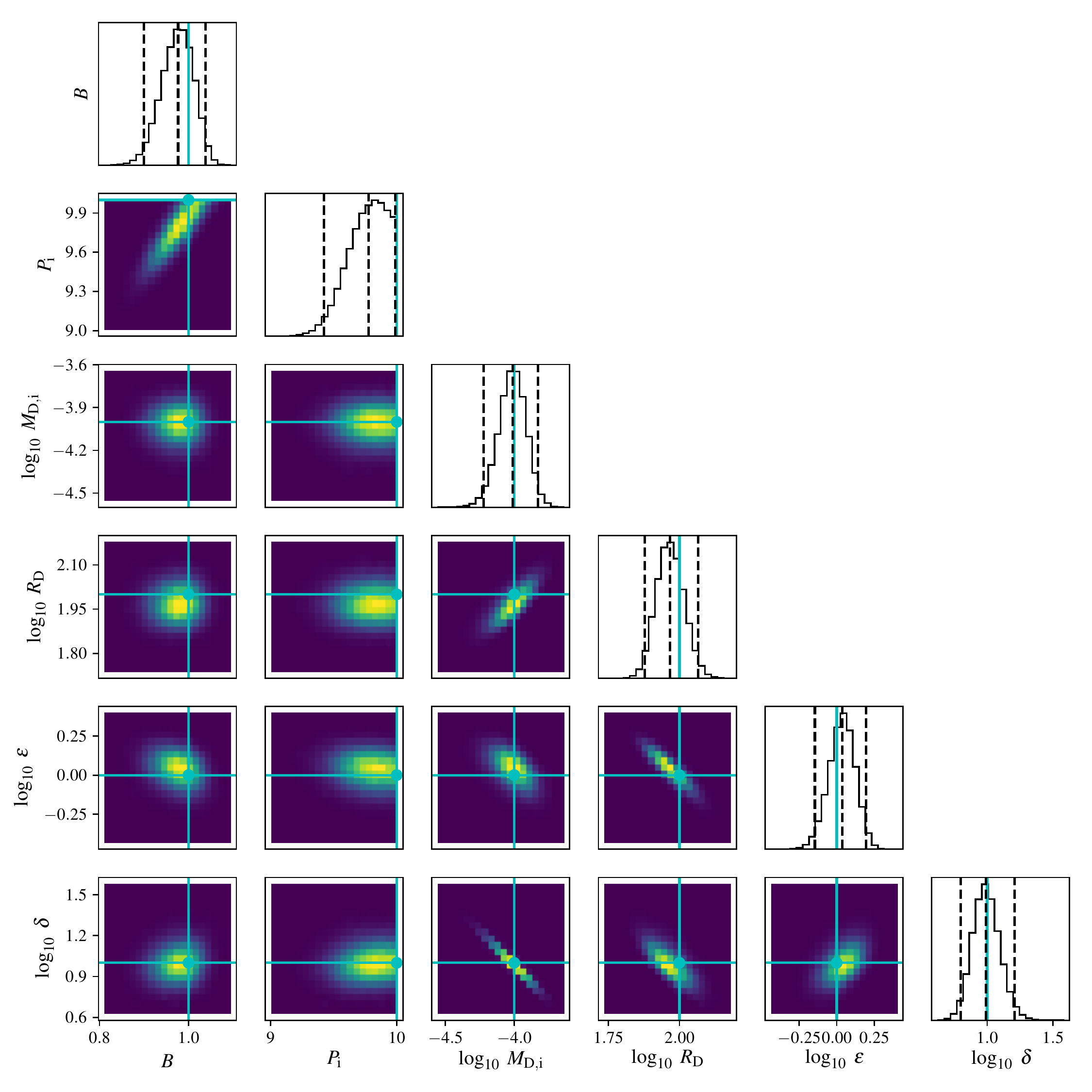}
\caption{2-D histograms showing the correlations between fitting parameters for a fit to a `stuttering' synthetic burst. The blue lines and points indicate the true values of the synthetic burst. The 1-D histograms show the sampled posterior distribution for each parameter. The dashed lines indicate the median and $\pm2\sigma$ values of the sampled posterior distribution and the blue line indicates the true value of the synthetic curve.}
\label{fig:stutt}
\end{figure*}

Previously in \citet{gibson17}, we recreated the $4$ morphologies of bursts the magnetic propeller model could produce as originally described by \citet{gompertz14}. These types are Type I `Humped', Type II `Classic', Type III `Sloped', and Type IV `Stuttering'. We chose a synthetic burst to represent each type and ran our MCMC algorithm, using $6$ fitting parameters, to test whether it could accurately reproduce the input values. Figures \ref{fig:humped}, \ref{fig:classic}, \ref{fig:sloped} and \ref{fig:stutt} are correlation plots generated from those MCMC runs. A strong and consistent correlation between $2$ parameters would indicate that they are degenerate.

While these plots reveal some strong correlations, notably $B-P_{\rm i}$, $\log\left(M_{\rm D,i}-\delta\right)$, $\log\left(\epsilon-\delta\right)$, the correlations change strength and shape for each burst type. In the `sloped' burst case (Fig.~\ref{fig:sloped}), the correlation between $B$ and $P_{\rm i}$ disappears completely because the dipole and propeller luminosity components are indistinguishable from one another in the light curve. Therefore, the parameters cannot always be degenerate with one another since the correlations change as the input parameters are varied and hence a treatment of these degeneracies is not required.

\begin{figure*}
\centering
\includegraphics[width=\textwidth]{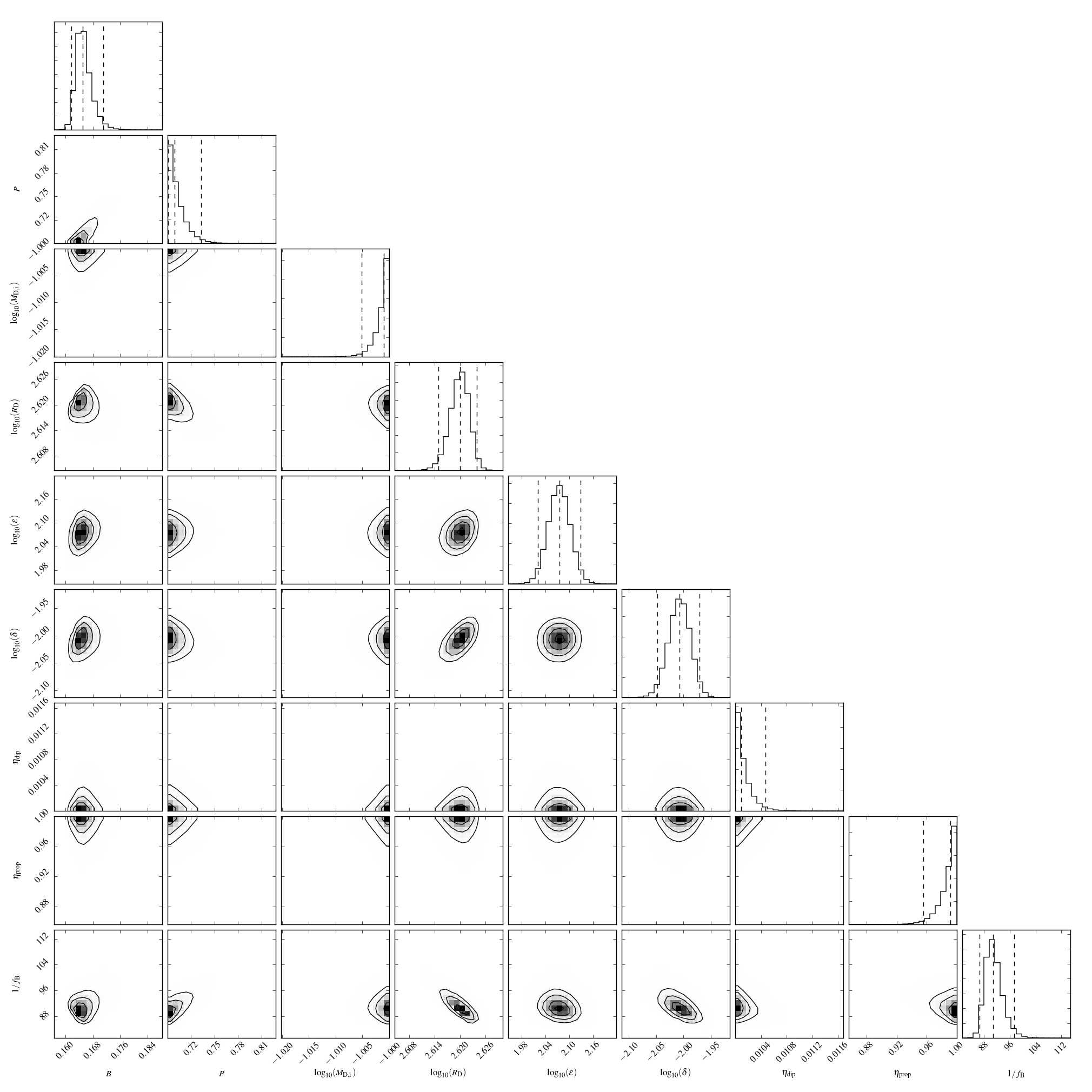}
\caption{2-D histograms showing the correlations between fitting parameters for the fit to GRB 060124. The 1-D histograms show the sampled posterior distribution for each parameter and the dashed lines indicate the median and $\pm2\sigma$ values corresponding to the values in Table \ref{tab:results}.}
\label{fig:060124}
\end{figure*}

\begin{figure*}
\centering
\includegraphics[width=\textwidth]{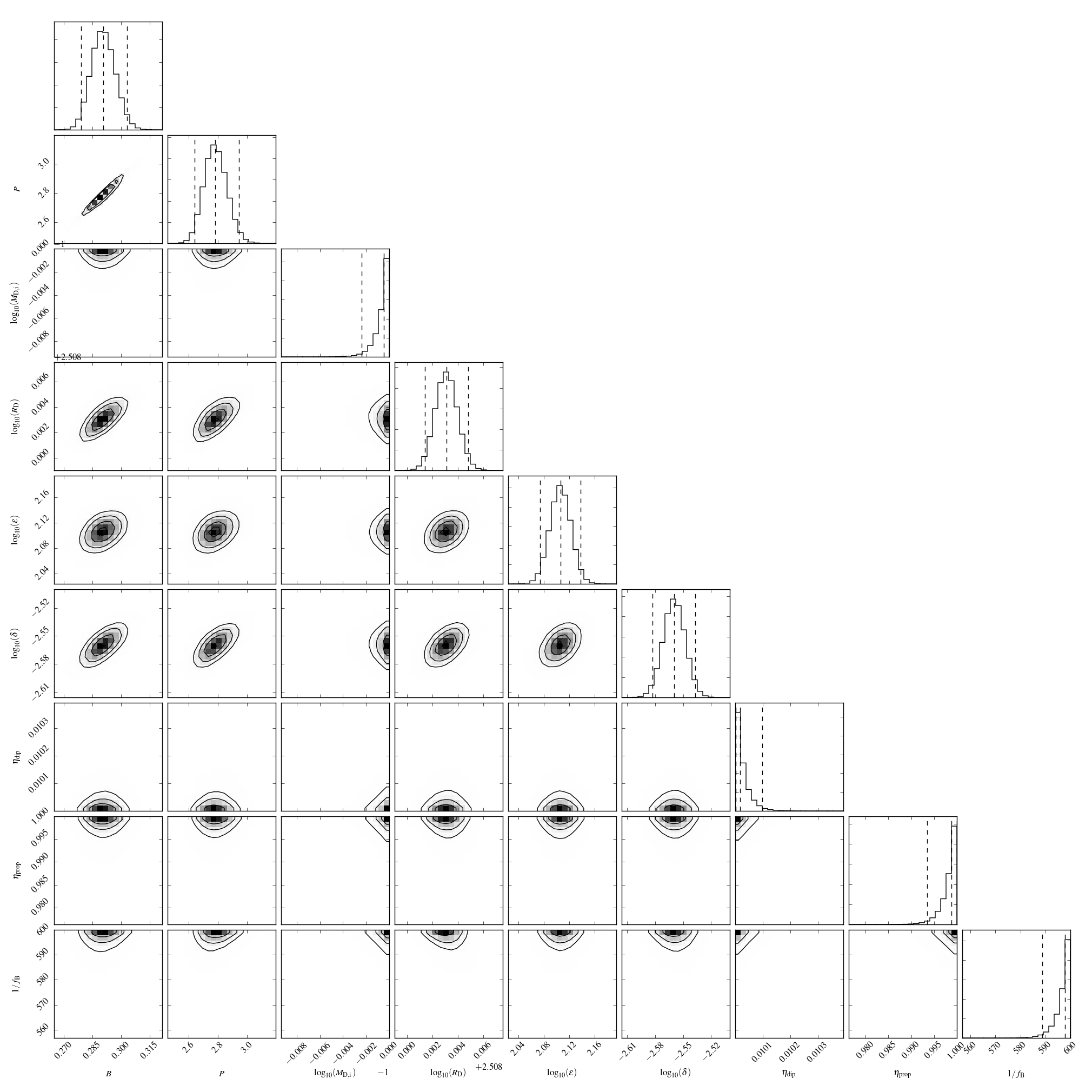}
\caption{2-D histograms showing the correlations between fitting parameters for the fit to GRB 060124. The 1-D histograms show the sampled posterior distribution for each parameter and the dashed lines indicate the median and $\pm2\sigma$ values corresponding to the values in Table \ref{tab:results}.}
\label{fig:121217A}
\end{figure*}

Similar plots for fits to GRBs 060124 and 121217A are presented in Figures \ref{fig:060124} and \ref{fig:121217A}, corresponding to the models and values in Fig.~\ref{fig:results} and Table \ref{tab:results} respectively. These plots further demonstrate the lack of requirement for a degeneracy treatment between fitting parameters since any correlations have mostly disappeared.

\bsp	
\label{lastpage}
\end{document}